\documentclass[conference]{IEEEtran}
\IEEEoverridecommandlockouts

\usepackage[letterpaper, left=1in, right=1in, bottom=1in, top=0.75in]{geometry}
\usepackage{algorithmic}
\usepackage{amsthm}
\usepackage{amsmath}
\usepackage{amssymb}
\usepackage{amsfonts}
\usepackage{balance}
\usepackage{caption}
\usepackage{cite}
\usepackage{color}
\usepackage{esint}
\usepackage{graphicx}
\usepackage{epstopdf}
\usepackage{subcaption}

\usepackage{epstopdf}

\usepackage{textcomp}
\def\BibTeX{{\rm B\kern-.05em{\sc i\kern-.025em b}\kern-.08em
    T\kern-.1667em\lower.7ex\hbox{E}\kern-.125emX}}
    
\theoremstyle{plain}
\newtheorem{thm}{\protect\theoremname}

\makeatother

\providecommand{\theoremname}{Theorem}

\theoremstyle{plain}

\makeatother

\providecommand{\lemmaname}{Lemma}

\theoremstyle{plain}

\makeatother

\providecommand{\propositionname}{Proposition}

\DeclareMathOperator{\erf}{erf}

\begin{document}

\title{Analytical Performance Evaluation of \\ THz Wireless Fiber Extenders
\thanks{This work has received funding from the European Commission’s Horizon 2020 research and innovation programme under grant agreement No. 761794.}
\thanks{This paper is the conference version of the paper~\cite{A:Analytical_Performance_Assessment_of_THz_Wireless_Systems}, which includes new insightful results.}
}

\author{\IEEEauthorblockN{ Alexandros--Apostolos A. Boulogeorgos, Evangelos N. Papasotiriou, and Angeliki Alexiou}
\IEEEauthorblockA{\textit{Department of Digital Systems,
University of Piraeus}, 
Piraeus, Greece \\
al.boulogeorgos@ieee.org, \{vangpapasot, alexiou\}@unipi.gr}
}

\maketitle

\begin{abstract}
This paper  presents the theoretical framework for the performance evaluation of terahertz (THz) wireless fiber extender in the presence of misalignment and multipath fading. In more detail, after providing  the appropriate system model that incorporates the different operation, design, and environmental parameters, such as the operation frequency, transceivers antenna gains, the level of misalignment as well as the stochastic behavior of the channel, we extract novel closed-form expressions for the ergodic capacity. 
These expressions are expected to be used as useful tools for the analysis and design of such systems. Moreover, several insightful scenarios are simulated. Their results highlight the importance of taking into account  the  impact of misalignment fading  when analyzing the performance of the THz wireless fiber extender. 
\end{abstract}


\vspace{-0.4cm}
\section{Introduction}\label{S:Intro}
\vspace{-0.2cm}

Terahertz (THz) fiber extenders have received considerable attention due to the unprecedented increase in the bandwidth and  ultra-high data rates that they offer in a cost-efficient and easy deployment manner~\cite{ref1_VTC,ref4_VTC,WP:Wireless_Thz_system_architecture_for_networks_beyond_5G} as well as the fact that they can contribute to bridging the  connectivity gaps that exist between the radio frequency (RF) access network and the fiber optic based backbone network~\cite{ref1_VTC}. This application scenario is expected to be employed in developing countries, where there may not be much of a fiber optic structure and hence to increase its reach and bandwidth to the last mile, without requiring a huge amount of economic resources to dig up the current brown-field~\cite{our_VTC_paper,C:UserAssociationInUltraDenseTHzNetworks}.

However, THz systems are subject to high path losses~\cite{WP:Wireless_Thz_system_architecture_for_networks_beyond_5G,C:ADistanceAndBWDependentAdaptiveModulationSchemeForTHzCommunications}, which originate from the band's high frequencies that cause interaction and energy absorptions by the molecules of the propagation medium~\cite{A:Propagation_modeling_for_wireless_communications_in_the_THz_band}. 
To understand the nature of these absorptions and to design countermeasures, a great amount of effort was put in modeling the THz channel particularities~\cite{ref6_VTC}, evaluating their impact on the system's performance~\cite{Capacity_and_throughput_analysis_of_nanoscale_machine} and proposing countermeasures~\cite{ref5_VTC,
Akkari2016,
A:Large_scale_antenna_systems_with_hybrid_analog_and_digital_beamforming_for_mmw_5G}. 
In particular, the  THz channel characteristics were studied in several works including~\cite{jornet2011,
EuCAP2018_cr_ver7,
our_PIMRC,
A:THz_Channel_model_and_link_budget_analysis_for_intrabody_nanoscale_communications,
A:THz_channel_modelling_of_wireless_ultra_compact_sensor_networks_using_em_waves,
A:Multi-ray_channel_modeling_and_wideband_characterixation_for_wireless_communications_in_the_THz_band}
and references therein. 
In more detail, in~\cite{jornet2011}, the authors presented a novel propagation model for THz nano-scale communications. Additionally, in~\cite{EuCAP2018_cr_ver7}, a simplified path-loss model for  the $275-400\text{ }\mathrm{GHz}$ band was introduced.
Furthermore, in~\cite{our_PIMRC}, a multi-ray THz propagation model was presented, while, in~\cite{A:THz_Channel_model_and_link_budget_analysis_for_intrabody_nanoscale_communications}, a propagation model for intra-body nano-scale communications was provided. 
Likewise, in~\cite{A:THz_channel_modelling_of_wireless_ultra_compact_sensor_networks_using_em_waves}, a path-loss model, which quantifies the total absorption loss assuming that air, natural gas and/or water are the components of the propagation medium, was reported, whereas, in~\cite{A:Multi-ray_channel_modeling_and_wideband_characterixation_for_wireless_communications_in_the_THz_band}, a multi-ray THz propagation model was~presented.

Although all the above  contributions  revealed the particularities of the THz medium, they neglected the impact of fading, which can be generated due to scattering on aerosols in the atmosphere~\cite{C:Frequency_domain_scattering_loss_in_THz_band}. 
On the contrary, in~\cite{paper_toyrkoi,
Alenka_2016,The_impact_of_antenna_directivties,shadowing_plants_2015}, the authors presented suitable stochastic models that are able to accommodate the multipath fading effect in the THz band. In particular, in~\cite{paper_toyrkoi}, the authors introduced a multi-path THz channel model, where the attenuation factor was modeled as a Rayleigh or Nakagami-$m$ distributions under the non-line-of-sight condition and as a Rician or Nakagami-$m$ distribution under the line-of-sight assumption. 
Moreover, in~\cite{Alenka_2016} and~\cite{The_impact_of_antenna_directivties}, the authors assumed Rician fading to accommodate the stochastic characteristics of the  THz communication channel. 
Likewise,  in~\cite{shadowing_plants_2015}, the authors used a log-normal shadowing path-loss model for THz nano-sensor~communications.
   
It is also widely known that another characteristic of THz systems that may cause  severe performance degradation is the misalignment between the transmitter (TX) and receiver (RX) antennas beams.
The impact of TX-RX beams misalignment in the THz link was discussed in~\cite{paper_toyrkoi,
The_impact_of_antenna_directivties} and ~\cite{A:Three_dimentional_end_to_end_modeling_and_analysis_for_graphene_enabled_THz_band_communications}. 
In more detail, in~\cite{paper_toyrkoi} and~\cite{A:Three_dimentional_end_to_end_modeling_and_analysis_for_graphene_enabled_THz_band_communications}, the authors assumed deterministic models to incorporate the impact of antennas' misalignment, while, in~\cite{The_impact_of_antenna_directivties}, it was considered to be a part of the shadowing effect. 
The disadvantage of these models is that they are unable to accommodate the stochastic characteristics of phenomena such as thermal expansion, dynamic wind loads and weak earthquakes, which result in the sway of high-rise buildings and cause vibrations of the transceivers antennas; in other words, the effect of misalignment  between the TX and RX~\cite{A:Average_capacity_for_heterodyne_FSO_communication_systems_over_g2_turbulence_channels_with_pointing_errors}.

To the best of the authors' knowledge, the  effect of misalignment  over fading channel in the THz band has not been addressed in the open  literature. Motivated by this,  this paper is focused on providing the  theoretical framework for the quantification of the performance of THz wireless fiber extenders. In particular, the technical contribution of this paper is outlined below:
\begin{itemize}
\item We establish an appropriate system  for the THz wireless fiber extenders, which takes into account the different design parameters, the channel characteristics, as well as their interactions. These parameters include the transmission range, the TX and RX antennas' gains, the degree of TX and RX misalignment, the transmission and noise~power.
\item In order to analytically evaluate the performance of the THz wireless extender, we extract novel closed-form expressions for the ergodic capacity.
\end{itemize}

\subsubsection*{Notations}
The operators $\mathbb{E}[\cdot]$ and $|\cdot|$ respectively denote the statistical expectation and the absolute value, whereas $\exp\left(x\right)$ and $\log_2\left(x\right)$ stand for the exponential function and the logarithmic function with base $2$.
Additionally, the operator $\ln\left(x\right)$ refers to the natural logarithm of $x$, while, the operator $\sqrt{x}$  returns the square root of $x$. Likewise, the set of the complex numbers is represented by~$\mathbb{C}$, while $CN\left(x, y\right)$ denotes a $x$-mean complex Gaussian process with variance $y$. 
 The upper and lower incomplete Gamma functions~\cite[eq. (8.350/2), (8.350/3)]{B:Gra_Ryz_Book} are respectively denoted by $\Gamma\left(\cdot, \cdot\right)$ and $\gamma\left(\cdot, \cdot\right)$, while the Gamma function is represented by~\cite[eq. (8.310)]{B:Gra_Ryz_Book}.
Finally, $\,_2F_1(\cdot,\cdot;\cdot; \cdot)$ and $G_{p, q}^{m, n}\left(x\left| \begin{array}{c} a_1, a_2, \cdots, a_{p} \\ b_{1}, b_2, \cdots, b_q\end{array}\right.\right)$ respectively stand for the Gauss hypergeometric function~\cite[eq. (4.1.1)]{B:Abramowitz} and the Meijer's G-function~\cite[eq. (9.301)]{B:Gra_Ryz_Book}, whereas $H_{p, q}^{m, n}\left[z \left|\begin{array}{c} (a_1, b_1), \cdots, (a_p, b_p) \\ (c_1, d_1), \cdots, (c_p, d_p) \end{array} \right. \right]$ is the Fox H-function~\cite[eq.  (8.3.1/1)]{B:Prudnikov_v3}.

\vspace{-0.4cm}
\section{System \& Channel Model}\label{sec:SM}


\begin{figure}
\centering\includegraphics[width=1\linewidth,trim=0 0 0 0,clip=false]{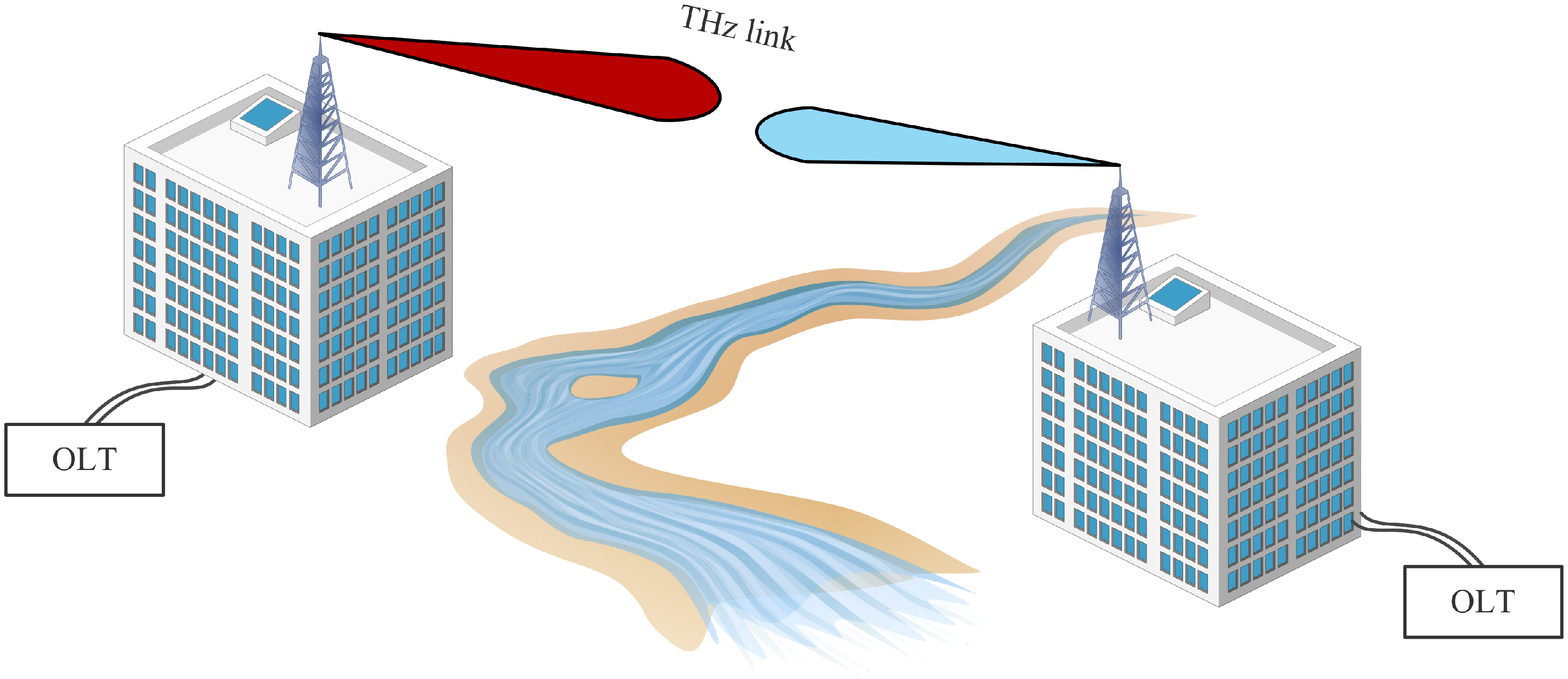}
\caption*{(a)}
\centering\includegraphics[width=0.6\linewidth,trim=0 0 0 0,clip=false]{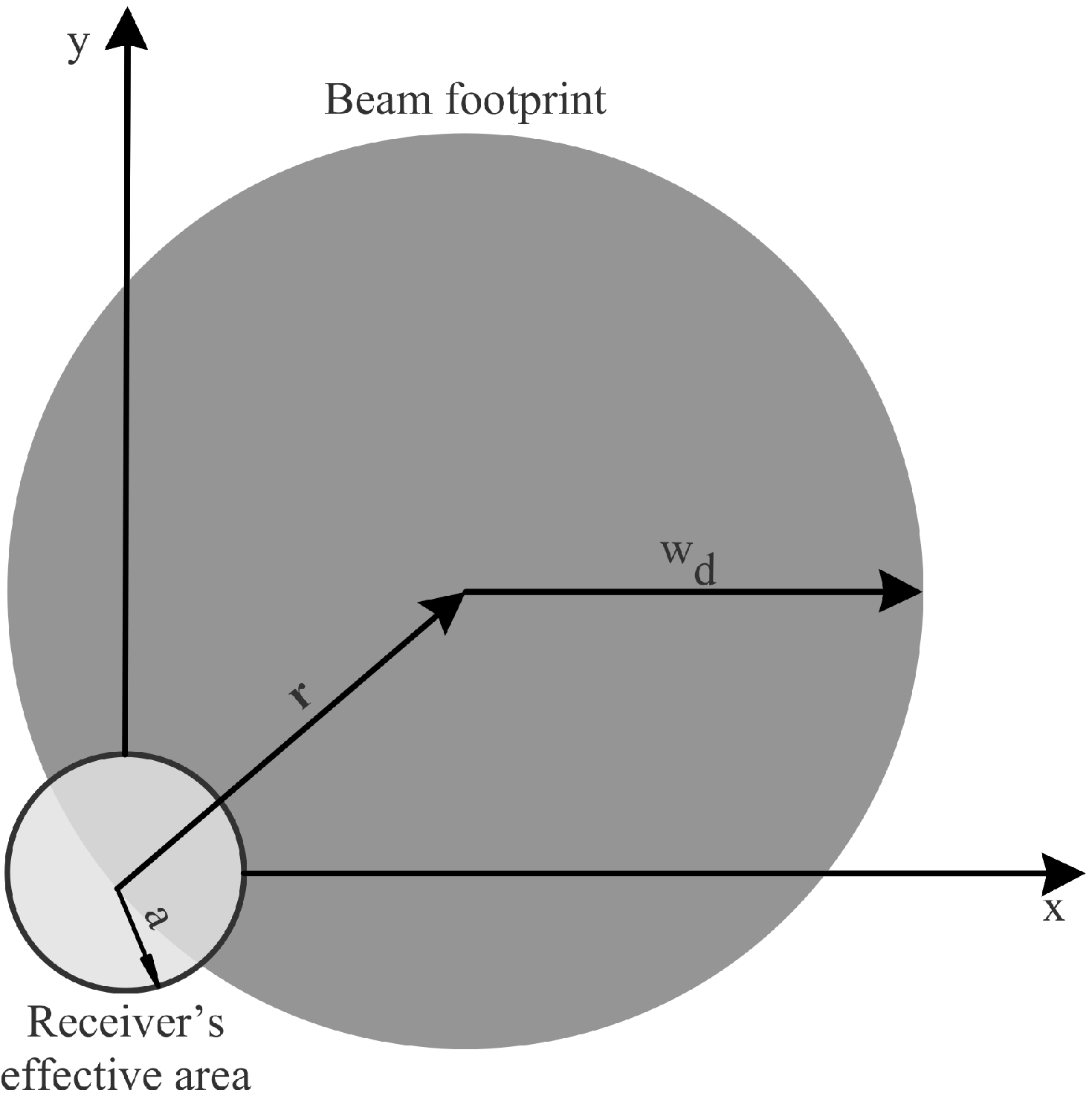}
\caption*{(b)}
\caption{(a) System model. OLT stands for the optical line terminal.
(b) RX's effective area and TX's beam footprint with misalignment on the RX's~plane.}
\label{fig:MF}
\end{figure}

As depicted in Fig.~\ref{fig:MF}.a, we consider a THz wireless fiber extender, in which  
the TX and RX  are equipped with highly directive antennas. The information signal, $x\in\mathbb{C}$, is conveyed over a flat fading wireless channel $h\in\mathbb{C}$ with additive noise $n\in\mathbb{C}$. Therefore, the baseband equivalent received signal can be expressed~as
\begin{align}
y = h x + n,
\label{Eq:y_ideal}
\end{align}  
where $h$, $x$ and $n$ are statistically independent. Likewise, $n$ is modeled as a complex zero-mean additive white Gaussian process with variance equals~$N_o$.


The channel coefficient, $h$, can be obtained~as 
\begin{align}
h =h_l h_p h_f,
\label{Eq:h}
\end{align}
where $h_l$, $h_p$ and $h_f$ respectively model the deterministic path gain, the misalignment fading, which results in pointing errors, and the  stochastic path gain.

The deteministic path gain  coefficient can be expressed~as
$h_l = h_{fl} h_{al},$
where $h_{fl}$ models the propagation gain and can be written~as
$h_{fl}=\dfrac{c\sqrt{G_t G_r}}{4 \pi f d}$,
where $G_\text{t}$ and $G_\text{r}$ respectively represent the antenna orientation dependent transmission and reception gains, $c$ stands for the speed of light, $f$ is the operating frequency and $d$ is the distance between the TX and the RX. 
Additionally, $h_{al}$ denotes the molecular absorption gain and can be evaluated as
$h_{al} = \exp\left(-\frac{1}{2}\kappa_{\alpha}(f) d\right)$,
where $\kappa_{\alpha}(f)$ denotes the absorption coefficient that describes the relative area per unit of volume, in which the molecules of the medium are capable of absorbing the electromagnetic wave energy, and can be evaluated as~\cite{EuCAP2018_cr_ver7}. 

As demonstrated in Fig.~\ref{fig:MF}.b, we assume that the RX has a circular detection beam of radius $\alpha$, covering an area $A$. 
Moreover, the TX has also a circular beam, which at distance $d$ has a radius $\rho$ that belongs in the interval $0 \leq \rho \leq w_d$, where $w_d$ is the maximum radius of the beam at distance $d$. Furthermore, both beams are considered on the positive Cartesian $x-y$ plane and $r$ is the pointing error expressed as the radial distance of the transmission and reception beams. Due to the symmetry of the beam shapes, $h_p$ depends only on the radial distance $r=|\emph{r}|$. Therefore, without loss of generality, it is assumed that the radial distance is located along the $x$ axis.
As a consequence and according to~\cite{Outage_capacity_FSO_2007},   the misalignment fading coefficient, $h_p$, which represents the fraction of the power collected by the RX, covering an area $A$ at distance $d$ can be approximated~as
\begin{align} 
h_p(r;d)\approx A_o \exp\left(-\dfrac{2 r^2}{w_{d_{eq}}^2}\right),
\label{eq:test1}
\end{align}
where $w_{d_{eq}}$ is the equivalent beam-width, whereas $A_o$ is the fraction of the collected power at $r=0$ and can be calculated~as  
$u = \frac{\sqrt{\pi} a}{\sqrt{2} w_{d}},$
with $a$ being the radius of the RX effective area and $w_d$ is the radius of the TX beam footprint at distance $d$. Additionally, the equivalent beam-width, $w_{eq}^2$, is related to $w_{d}^2$ through
$w_{eq}^2 = w_{d}^{2} \frac{\sqrt{\pi} \erf\left(u\right)}{2 u \exp\left(-u^2\right)}.$

By considering independent identical Gaussian distributions for the elevation and horizontal displacement~\cite{Outage_capacity_FSO_2007,arnon2003}, the radial displacement at the RX follows a Rayleigh distribution with probability density function (PDF), which can be expressed as
\begin{align} 
f_r(r)=\frac{r}{\sigma^{2}_{s}} \exp\left(\frac{r^{2}}{2\sigma^{2}_{s}}\right),
\label{eq:test2} 
\end{align} 
where $\sigma^{2}_s$ is the variance of the pointing error displacement at the RX. 
By combining~\eqref{eq:test1} and~\eqref{eq:test2}, the PDF of $|h_p|$ can be expressed as~\cite{Outage_capacity_FSO_2007}
\begin{align}
f_{h_{p}}(x)=\frac{\xi}{A^{\xi}_{o}}x^{\xi-1}, \text{ } 0 \leq x \leq A_o,
\label{Eq:f_hp}
\end{align}
where 
$\xi=\sqrt{\frac{w_{eq}}{2\sigma_s}}$
is the ratio between the equivalent beam width radius at the RX. 
Note that this model was extensively used in several studies in free space optical systems (see e.g.,~\cite{sandalidis2009,
A:Performance_Analysis_of_Relay_Assisted_All_Optical_FSO_Networks_Over_Strong_Atmospheric_Turbulence_Channels_With_Pointing_Errors} and references therein).  

The stochastic path gain coefficient, $|h_f|$, is assumed to follow an  $\alpha-\mu$ distribution~\cite{alpha_mu_Yacoub_2007}, with PDF given~by
\begin{align}
f_{h_{f}}(x)=\dfrac{\alpha \mu^{\mu} }{\hat{h}_{f}^{\alpha \mu} \Gamma(\mu)} x^{\alpha \mu - 1} \exp\left(-\mu \dfrac{x^{\alpha}}{\hat{h}_f^{\alpha}}\right),
\label{Eq:f_hf}
\end{align}
where $\alpha > 0$  is a fading parameter, $\mu$ is the normalized variance of the fading channel envelope, and $\hat{h}_{f}$ is the $\alpha$-root mean value of the fading channel envelop. Note that this distribution is a general form for many well-known distributions, such as Rayleigh ($\alpha=2, \mu=1$), Nakagami-$m$ ($\alpha=2$ and $\mu$ is the fading parameter), Weibull ($\mu=1$ and $\alpha$ is the fading parameter), etc.~\cite{A:On_physical_layer_security_over_generalized_gamma_fading_channels}.  


\section{Performance Analysis}\label{sec:PA}

Theorem~\ref{Thm:Ergodic_Capacity_Ideal_Exact} returns a novel closed-form expression for the evaluation of the ergodic capacity.

\begin{thm}\label{Thm:Ergodic_Capacity_Ideal_Exact}
The ergodic capacity can be analytically expressed as in~\eqref{Eq:Ergodic_Capacity_Ideal_Exact}, given at the top of the next page.
\begin{figure*}
\begin{align}
{C}&= \frac{\Theta \Delta^{-\frac{\Lambda+1}{2}}}{\ln\left(2\right)} 
H_{3, 4}^{4, 1}\left[\frac{{X}}{\Delta^{\frac{\alpha}{2}}}\left|
\begin{array}{c}
\left( -\frac{\Lambda+1}{2}, \frac{\alpha}{2}\right), \left(1-\frac{\Lambda+1}{2},\frac{\alpha}{2}\right), \left(1, 1\right) \\
\left(0, 1\right), \left({\Phi}, 1\right), \left(-\frac{\Lambda+1}{2},\frac{\alpha}{2}\right), \left(-\frac{\Lambda+1}{2}, \frac{\alpha}{2}\right)
\end{array}
 \right.\right] 
 \label{Eq:Ergodic_Capacity_Ideal_Exact}
\end{align}
\hrulefill
\end{figure*}
In~\eqref{Eq:Ergodic_Capacity_Ideal_Exact},
$\Delta=\frac{|h_l|^{2} P}{N_{o}},$
$\Lambda=\xi -1,$ 
${\Phi}= \frac{\alpha \mu - \xi}{\alpha},$
${X}=  \frac{\mu}{\widehat{h}_f^{\alpha} A_0^{a}}$
and 
$\Theta =\xi A_0^{-\xi}  \frac{\mu^{\mu-\frac{\mu\alpha-\xi}{\alpha}}}{\hat{h}_f^{\alpha}\Gamma\left(\mu\right)}.$
\end{thm}
\begin{IEEEproof}
Based on~\eqref{Eq:y_ideal}, the instantaneous SNR of the received signal can be obtained~as
\begin{align}
\gamma= \frac{|h|^{2}P}{N_{o}}.
\label{Eq:SNR}
\end{align}
Hence, the ergodic capacity can be defined~as
$C = \mathbb{E}\left[\log_2\left(1+\gamma_i\right)\right],$
or
\begin{align}
C = \mathbb{E}\left[\log_2\left(1+\Delta|h_{fp}|^2\right)\right],
\end{align}
which can be evaluated~as
\begin{align}
C = \int_{0}^{\infty} \log_2\left(1+\Delta x^2\right) f_{|h_{fp}|}(x) \mathrm{d}x,
\label{Eq:Ergodic_capacity_ideal_step1}
\end{align}
where $f_{|h_{fp}|}(x)$ stands for the probability density function of the random variable $h_{fp}=h_f h_p$ and can be evaluated, according to~\cite{papoulis},~as
\begin{align}
f_{|h_{fp}|}(x) = 
\int_{0}^{A_o} \frac{1}{y} f_{|h_f|}\left(\frac{x}{y}\right) 
f_{|h_m|}(y) \mathrm{d}y.
\label{Eq:f_hfp_definition}
\end{align} 
By substituting~\eqref{Eq:f_hp} and~\eqref{Eq:f_hf} into~\eqref{Eq:f_hfp_definition}, the PDF of $|h_{fp}|$ can be equivalently expressed~as
\begin{align}
f_{|h_{fp}|}(x) = \frac{\alpha \mu^\mu x^{\alpha \mu -1}}{\left(\hat{h}_f\right)^{\alpha \mu}\Gamma(\mu)} \frac{\xi}{A_o^{\xi}}\mathcal{I}(x),
\label{Eq:f_hfp2} 
\end{align}
where
\begin{align}
\mathcal{I}(x) = \int_{0}^{A_o}y^{\xi-\alpha\mu-1} \exp\left(-\mu\frac{x^{\alpha} y^{-\alpha}}{\hat{h}^{\alpha}}\right) \mathrm{d}y.
\label{Eq:Ix}
\end{align}
By employing~\cite[eq. (8/350/1)]{B:Gra_Ryz_Book},~\eqref{Eq:f_hfp2} can be written~as 
\begin{align}
f_{|h_{fp}|}(x) = \xi 
A_0^{-\xi}  &
\frac{\mu^{\frac{\xi}{\alpha}}}{\hat{h}_f^{\alpha}\Gamma\left(\mu\right)} 
x^{\xi-1} 
\nonumber \\ & \times
\Gamma\left(
\frac{\alpha \mu - \xi}{\alpha}, 
\mu \frac{x^{\alpha}}{\hat{h}_f} A_0^{-\alpha}
\right).
\label{Eq:f_{|h_{fp}|}}
\end{align}

Next, by substituting~\eqref{Eq:f_{|h_{fp}|}} into~\eqref{Eq:Ergodic_capacity_ideal_step1}, the ergodic capacity can be expressed~as
\begin{align}
C = \frac{\Theta}{\ln\left(2\right)} \int_{0}^{\infty} x^{\Lambda} \ln\left(1+\Delta x^2\right)  \Gamma\left(\Phi, X x^{\alpha}\right) \mathrm{d}x,
\label{Eq:Ergodic_capacity_ideal_step2}
\end{align}
or, by using~\cite[eq. (15.1.1)]{B:Abramowitz},~as
\begin{align}
C \hspace{-0.1cm}=\hspace{-0.1cm} \frac{\Theta}{\ln\left(2\right)} \int_{0}^{\infty}\hspace{-0.4cm} x^{\Lambda+1} \,_2F_1\left(1, 1; 2; - x\right) \Gamma\left(\Phi, X x^{\alpha}\right) \mathrm{d}x.
\label{Eq:Ergodic_capacity_ideal_step3}
\end{align}
Moreover, by employing~\cite[eq. (11)]{C:The_algorithm_for_calcuating_integrals_of_hypergeometric_type_functions_and_its_realization_in_Reduce_system} and~\cite[eq. (9.34/7)]{B:Gra_Ryz_Book},~\eqref{Eq:Ergodic_capacity_ideal_step3} can be written~as
\begin{align}
C = \frac{\Theta}{\ln\left(2\right)} \int_{0}^{\infty} &x^{\Lambda}  G_{2, 2}^{1, 2}\left( \Delta x^2 \left| \begin{array}{c} 1, 1 \\ 1, 0\end{array}\right.\right)
\nonumber \\ & \times
G_{1, 2}^{2, 0}\left( X x^{\alpha}\left| \begin{array}{c} 1 \\ 0, \Phi \end{array}\right. \right)
\mathrm{d}x,
\label{Eq:Ergodic_capacity_ideal_step4}
\end{align}
which, by setting $z={x^{2}}$, can be expressed~as
\begin{align}
C = \frac{\Theta}{\ln\left(2\right)} \int_{0}^{\infty} &x^{\frac{\Lambda+1}{2}-1}  G_{2, 2}^{1, 2}\left( \Delta z \left| \begin{array}{c} 1, 1 \\ 1, 0\end{array}\right.\right)
\nonumber \\ & \times
G_{1, 2}^{2, 0}\left( X x^{\frac{\alpha}{2}}\left| \begin{array}{c} 1 \\ 0, \Phi \end{array}\right. \right)
\mathrm{d}z.
\label{Eq:Ergodic_capacity_ideal_step5}
\end{align}
Finally, by using~\cite{site:Wolfram} into~\eqref{Eq:Ergodic_capacity_ideal_step5}, the ergodic capacity can be rewritten as~\eqref{Eq:Ergodic_Capacity_Ideal_Exact}.
This concludes the proof. 
\end{IEEEproof}

\vspace{-0.3cm}
\section{Results and Discussion}\label{sec:Results_Discussion}
\vspace{-0.1cm}
In this section, we present the joint effect of the deterministic and stochastic path-gain, i.e., misalignment and multipath fading, components in the ergodic capacity of the THz wireless fiber extender by illustrating analytical and  simulation results. Unless otherwise is stated,  it is assumed that $G_t=G_r=55\text{ }\mathrm{dBi}$\footnote{According to~\cite{A:Wireless_Sub_THz_Communication_System_With_High_Data_Rate} and~\cite{A:On_millimeter_wave_and_THz_mobile_radio_channel_for_smart_rail_mobility}, this antenna gain can be practically achieved by employing high-gain Cassegrain antennas.}, $\alpha=2$, and $\mu=4$.
Moreover, standard environmental conditions, i.e., $\phi=50\%$, $p=101325\text{ }\mathrm{Pa}$, and $T=296\text{ }^o\mathrm{K}$, are assumed. 
Finally, note that all the analytical results were verified via respective Monte-Carlo simulations. 




\begin{figure}
\centering\includegraphics[width=0.83\linewidth,trim=0 0 0 0,clip=false]{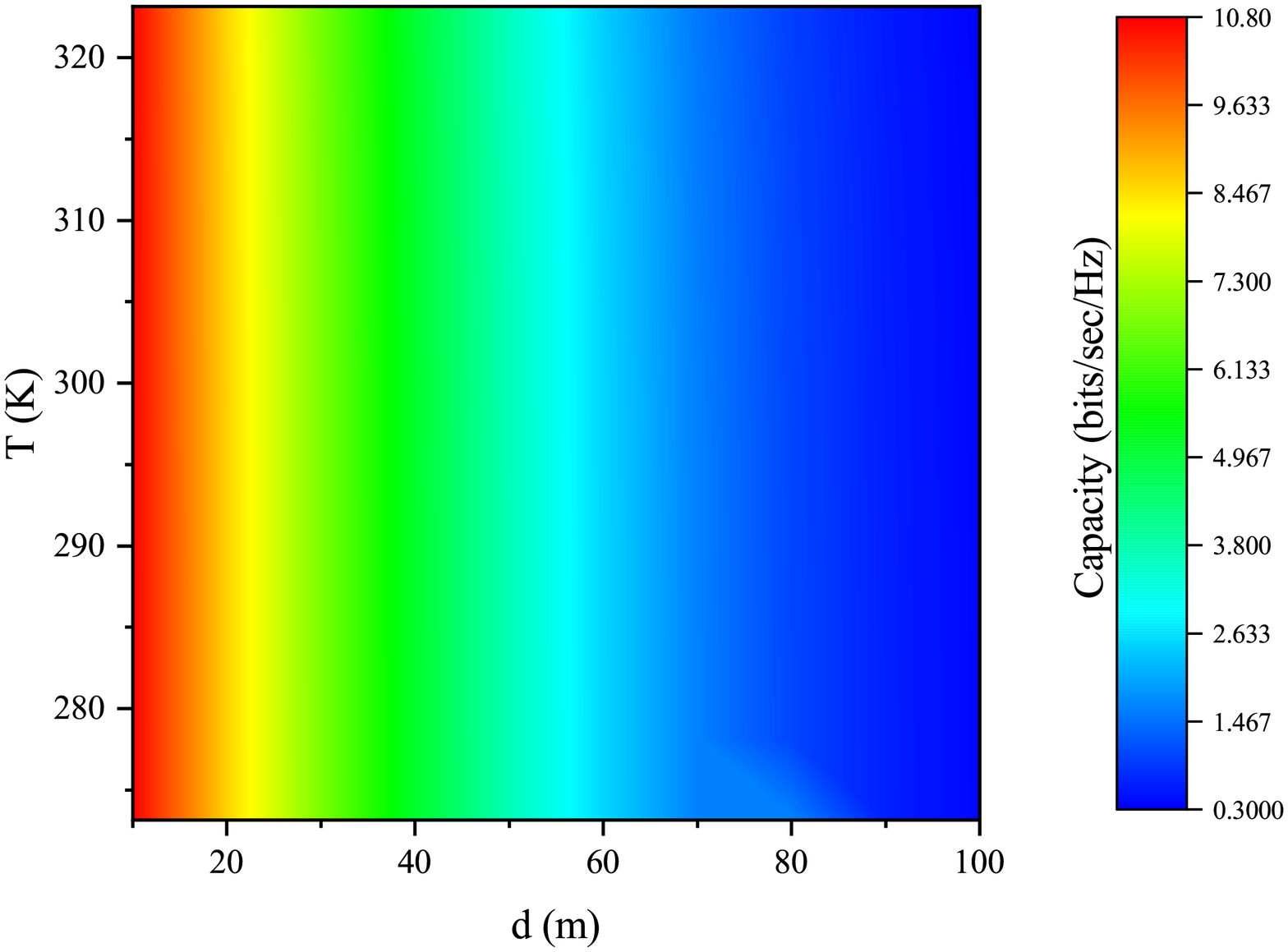}
\caption{Capacity vs distance and Temperature, for $f=300\text{ }\mathrm{GHz}$, $\frac{P}{N_o}=25\text{ }\mathrm{dB}$, $\sigma_s=0.01\text{ }\mathrm{m}$}
\label{fig:EC_vs_rh_T}
\end{figure}

Fig.~\ref{fig:EC_vs_rh_T} shows the ergodic capacity as a function of the transmission range and atmospheric temperature, assuming $f=300\text{ }\mathrm{GHz}$,  $\frac{P}{N_o}=25\text{ }\mathrm{dB}$ and $\sigma_s=0.01\text{ }\mathrm{m}$. 
We observe that, for a fixed temperature, as the transmission range increases, the deterministic path-loss also increases; hence, the ergodic capacity decreases. 
Similarly, for a given transmission range, as the temperature increases, the molecular absorption loss also increases; therefore, the ergodic capacity degrades. 
Finally, it is evident that the impact of propagation loss is  more severe compared to the one of the molecular absorption, since the ergodic capacity degradation due to a transmission distance increase is considerably more detrimental than the corresponding degradation due to a temperature increase. 

\begin{figure}
\centering\includegraphics[width=0.83\linewidth,trim=0 0 0 0,clip=false]{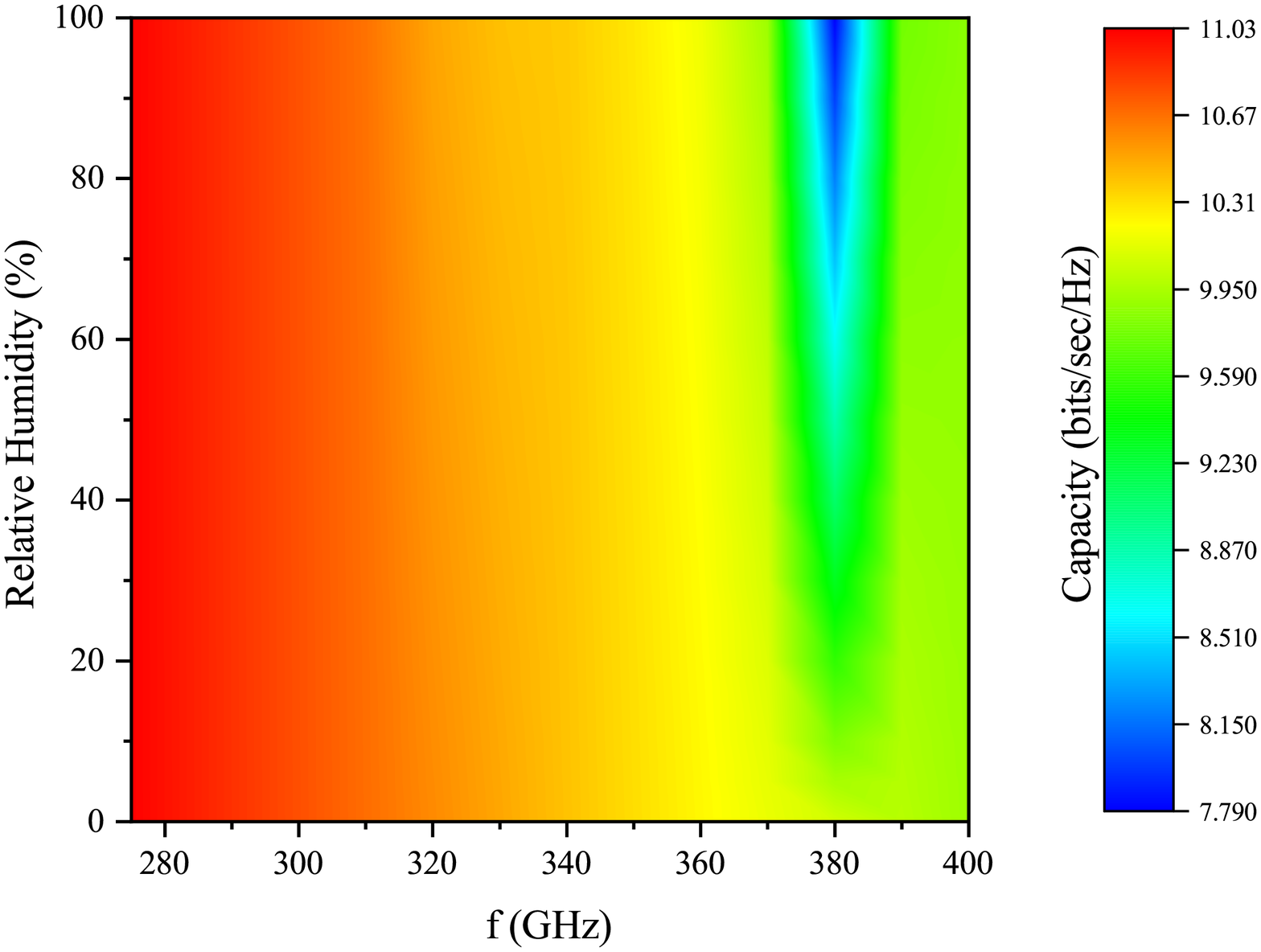}
\caption{Capacity vs frequency and relative humidity, for  $d=10\text{ }\mathrm{m}$,  $\frac{P}{N_o}=25\text{ }\mathrm{dB}$, $\sigma_s=0.01\text{ }\mathrm{m}$.}
\label{fig:EC_vs_rh_f}
\end{figure}

Fig.~\ref{fig:EC_vs_rh_f} demonstrates the ergodic capacity as a function of frequency and relative humidity, assuming  $d=10\text{ }\mathrm{m}$,  $\frac{P}{N_o}=25\text{ }\mathrm{dB}$ and $\sigma_s=0.01\text{ }\mathrm{m}$. From this figure, it is evident that the effect of  humidity on the ergodic capacity depends on the operation frequency. 
In more detail, we observe that up to $320\text{ }\mathrm{GHz}$ the effect of relative humidity alteration is relatively low, whereas in the $380\text{ }\mathrm{GHz}$ region, it is detrimental.
For example, in $300\text{ }\mathrm{GHz}$, a relative humidity alteration from $30\%$ to $70\%$ results to a $0.03\%$ ergodic capacity degradation, whereas the same alteration in the $380\text{ }\mathrm{GHz}$ causes a corresponding $9.7\%$ decrease.  
This indicates the importance of taking into account the variation of the environmental conditions in a geographical area, when selecting the operation frequency.


\begin{figure}
\centering\includegraphics[width=0.83\linewidth,trim=0 0 0 0,clip=false]{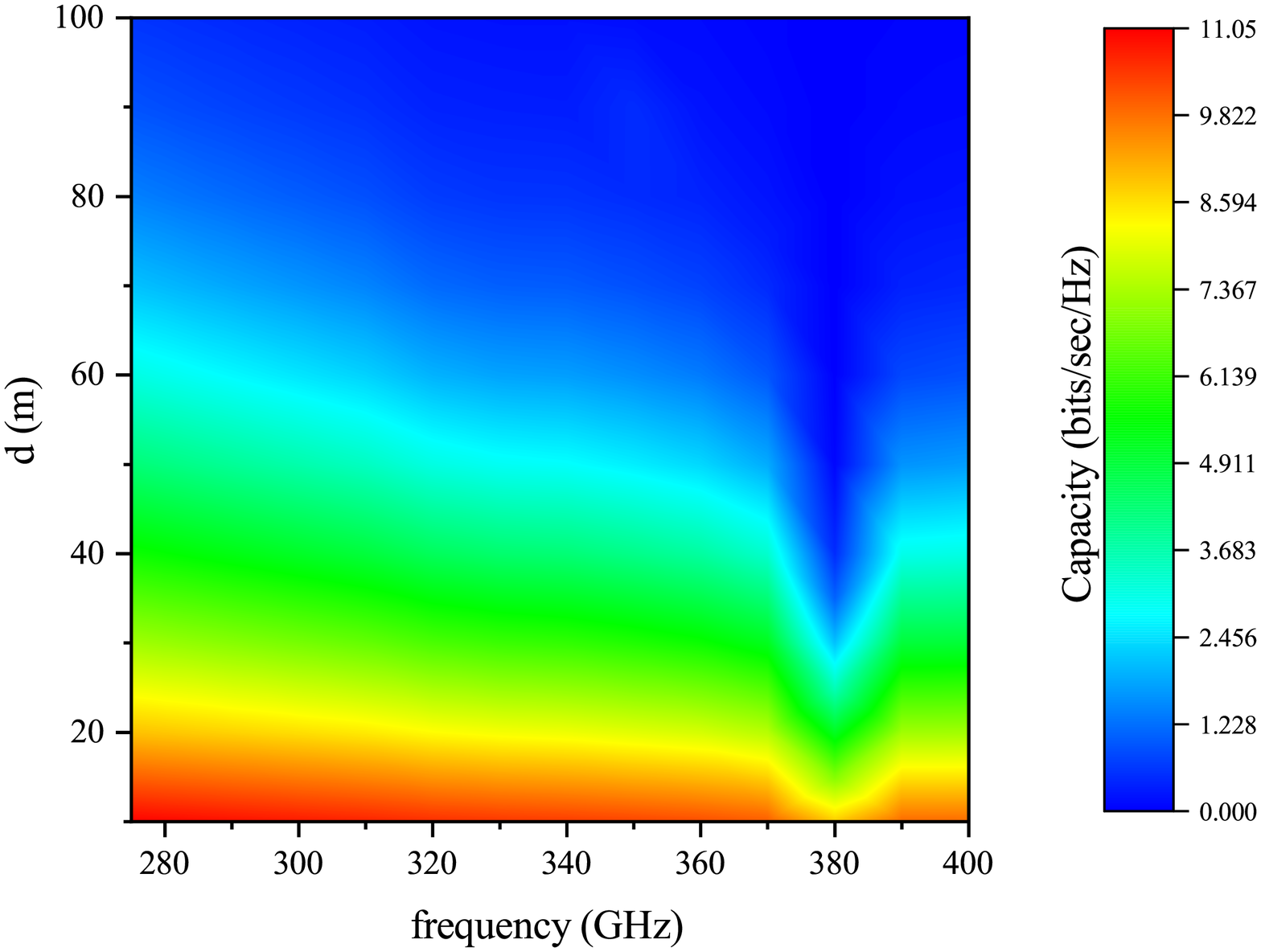}
\caption{Capacity vs  frequency and distance, $\frac{P}{N_o}=25\text{ }\mathrm{dB}$ and $\sigma_s=0.01\text{ }\mathrm{m}$.}
\label{fig:Capacity_vs_f_vs_d}
\end{figure}

Fig.~\ref{fig:Capacity_vs_f_vs_d} illustrates the ergodic capacity as function of frequency and distance, assuming $\frac{P}{N_o}=25\text{ }\mathrm{dB}$ and $\sigma_s=0.01\text{ }\mathrm{m}$. For a given operation frequency, as the distance increases the total path-loss also increases; hence, the ergodic capacity significantly decreases. Furthermore, for a fixed transmission distance, as the frequency increases, the path-loss drastically increases; hence, the  ergodic capacity  detrimentally decreases. For example, for $f=340\text{ }\mathrm{GHz}$ increasing the transmission distance from $20\text{ }\mathrm{m}$ to $60\text{ }\mathrm{m}$ causes an ergodic capacity degradation of approximately $78.6\%$. Similarly, for $d=20\text{ }\mathrm{m}$ and a frequency alteration from $300\text{ }\mathrm{GHz}$ to $400\text{ }\mathrm{GHz}$ results to an ergodic capacity decrease by about $15.7\%$.

\begin{figure}
\centering\includegraphics[width=0.68\linewidth,trim=0 0 0 0,clip=false]{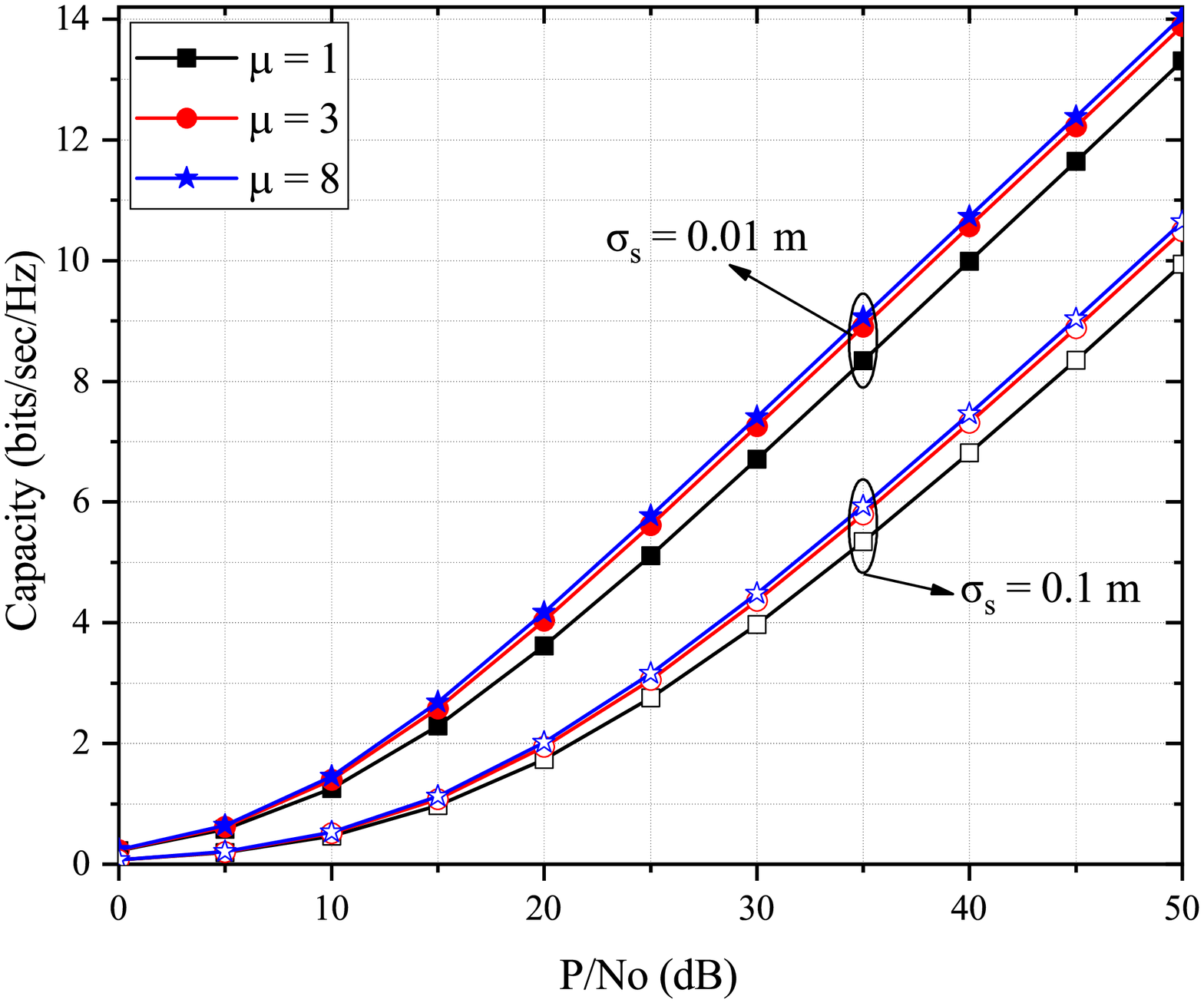}
\caption{Capacity vs  $\frac{P}{N_o}$ for different values of $\mu$ and $\sigma_s$, for $f=275\text{ }\mathrm{GHz}$ and $d=40\text{ }\mathrm{m}$.}
\label{fig:Capacity_vs_P_No_diff_mu_sigma_s}
\end{figure}

\begin{figure}
\centering\includegraphics[width=0.68\linewidth,trim=0 0 0 0,clip=false]{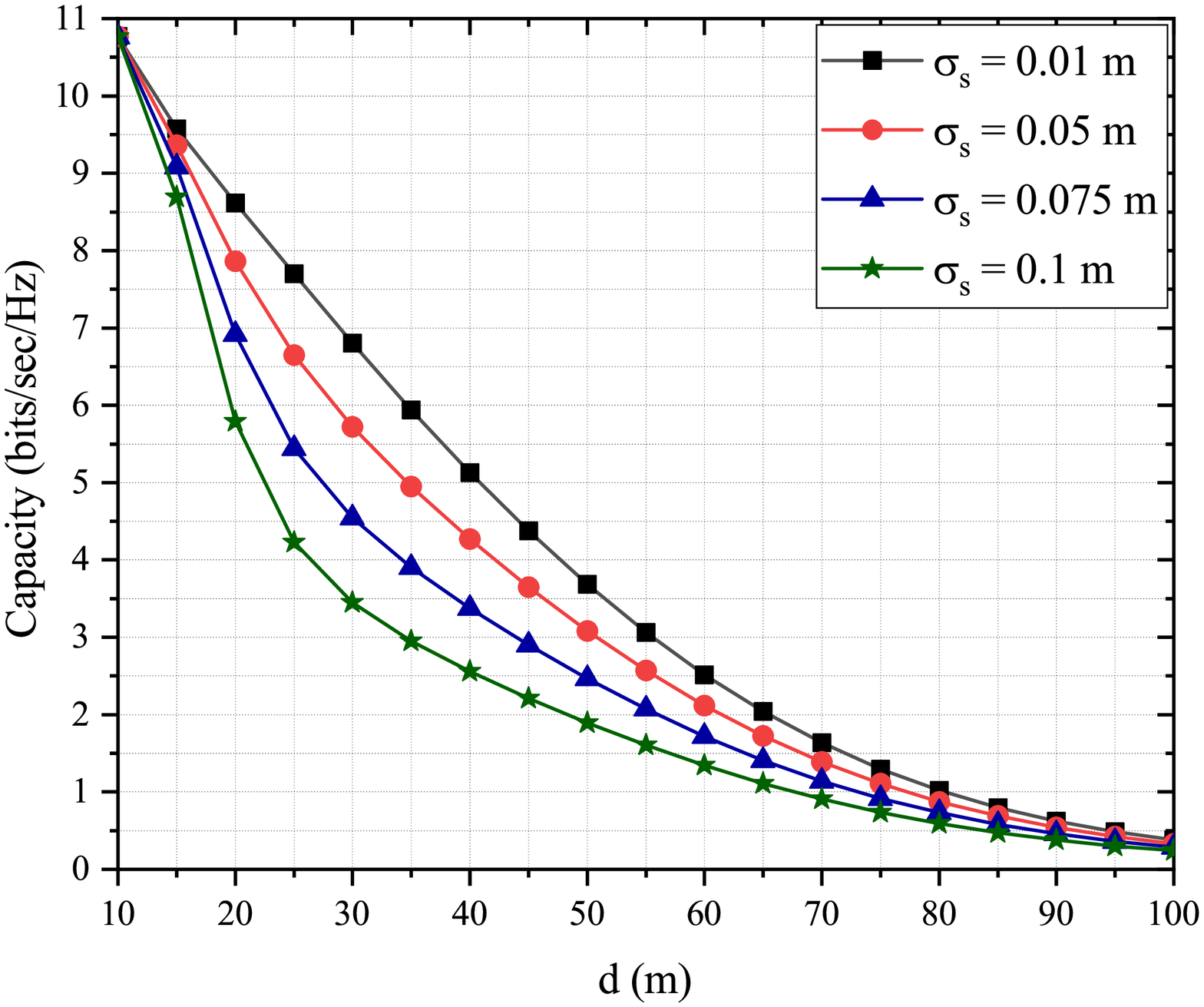}
\caption{Capacity vs  distance for different values of $\sigma_s$, $f=300\text{ }\mathrm{GHz}$ and $\frac{P}{N_o}=25\text{ }\mathrm{dB}$.}
\label{fig:Capacity_vs_d_sigma_s}
\end{figure}

Fig.~\ref{fig:Capacity_vs_P_No_diff_mu_sigma_s} depicts the ergodic capacity as a function of $\frac{P}{N_o}$, for different values of $\mu$ and $\sigma_s$, assuming $f=275\text{ }\mathrm{GHz}$ and $d=40\text{ }\mathrm{m}$. The markers and lines indicate the fading parameter $\mu$, specifically the black ones stand for $\mu=1$, the red ones for $\mu=3$ and the blue ones for $\mu=8$. Also, the color filled markers denote the ergodic capacity calculated with $\sigma_s=0.01\text{ }\mathrm{m}$, while the non-filled markers represent the results with $\sigma_s=0.1\text{ }\mathrm{m}$. Note that the ergodic capacity of the Rayleigh multipath fading $\left(\mu=1 \right) $ is provided in this figure to serve as benchmark for the worst case scenario. For a given $\frac{P}{N_o}$ and $\sigma_s$, as the fading parameter $\mu$ increases the effect multipath fading decreases; therefore, the overall SNR improves yielding greater capacity. For example, for $\frac{P}{N_o}=40\text{ }\mathrm{dB}$ and $\sigma_s=0.01\text{ }\mathrm{m}$ changing $\mu$ from $1$ to $3$ and from $1$ to $8$, causes a  $5.8\%$ and $7.4\%$  ergodic capacity increase,  respectively, while for the same $\frac{P}{N_o}$ and $\sigma_s=0.1\text{ }\mathrm{m}$, the same $\mu$ alterations lead to an ergodic capacity increase of $7.3\%$ and $9.5\%$, respectively. Furthermore, for a given $\frac{P}{N_o}$ and $\mu$, the increase of jitter variance significantly deteriorates the achievable ergodic capacity. For instance, for $\frac{P}{N_o}=30\textbf{ }\mathrm{dB}$ and $\mu=3$, the increase of $\sigma_s$ from $0.01\textbf{ }\mathrm{m}$ to $0.1\textbf{ }\mathrm{m}$ results to an ergodic capacity degradation of $40\%$. Finally, this figure reveals that the impact of misalignment fading is somewhat more detrimental compared to the one of multipath fading.

Fig.~\ref{fig:Capacity_vs_d_sigma_s} presents the ergodic capacity as a function of the transmission distance, for different values of $\sigma_s$, assuming $\frac{P}{N_o}=25\text{ }\mathrm{dB}$ and $f=300\text{ }\mathrm{GHz}$. As expected, for a given $\sigma_s$, the increase of the transmission distance causes detrimental increase of the path-loss; as a consequence, the achievable ergodic capacity severely degrades. For example, for $\sigma_s=0.05\text{ }\mathrm{m}$, a transmission distance change from $20\text{ }\mathrm{m}$ to $50\text{ }\mathrm{m}$ reduces the achieved ergodic capacity by $60.8\%$. Furthermore, for a fixed transmission distance, the increase of $\sigma_s$ drastically deteriorates the ergodic capacity. For instance, for  $d=40\text{ }\mathrm{m}$, a $\sigma_s$ alteration from $0.01\text{ }\mathrm{m}$ to $0.05\text{ }\mathrm{m}$, and $0.1\text{ }\mathrm{m}$ results to an ergodic capacity degradation of $16.8\%$, while a $\sigma_s$ change from $0.01\text{ }\mathrm{m}$ to $0.075\text{ }\mathrm{m}$ in a $34.2\%$ ergodic capacity degradation. This indicates the importance of taking into account the level of antennas' misalignment, when evaluating the ergodic capacity performance of the THz wireless fiber extender.

\section{Conclusions}\label{sec:Conclusions}

We studied the performance of wireless THz fiber extenders, when the transceivers antennas are not fully-aligned. In particular, by assuming $\alpha-\mu$ fading and Gaussian distributions for the elevation and horizontal displacement, we provided the analytical framework for evaluating the ergodic capacity.
Our results illustrated the degradation due to the  effect of misalignment fading  on the ergodic capacity  of the THz wireless fiber extender.
It was also revealed that the impact of misalignment fading is somewhat more detrimental compared to the one of the multipath fading. 
Finally,  the importance of accurate misalignment characterization in the realistic performance analysis and design of THz wireless fiber extenders was highlighted.

\balance
\bibliographystyle{IEEEtran}
\bibliography{IEEEabrv,References}
\balance

\end{document}